\documentclass[10pt]{article}
\usepackage{a4wide}                      % to get a wider page margen
\usepackage{epsf}                        % to make figures
\usepackage{latexsym}                    % to make fancy symbols
\usepackage{amsfonts}                    % to make fancy symbols
\usepackage{amssymb}                     % to make fancy symbols
\newcommand{\sect}[1]{\setcounter{equation}{0}\section{#1}}

\def\be{\begin{equation}}
\def\ee{\end{equation}}
\def\bea{\begin{eqnarray}}
\def\eea{\end{eqnarray}}

\def\nnw{\nonumber \\ [.2cm]}
\def\vsp#1{\vspace*{#1}}
\def\hsp#1{\hspace*{#1}}

\def\part{\partial}
\def\tfrac#1#2{{\textstyle{\frac{#1}{#2}}}}
\def\half{\tfrac{1}{2}}
\def\x{\times}

\def\Tr{\mbox{Tr}}
\def\Str{\mbox{STr}}
\def\STr{\mbox{STr}}

\def\incl{\mbox{i}}
 
\def\cF{{\cal F}} 
\def\cG{{\cal G}}

\def\Ortin{Ort{\'\i}n}
\def\RodrGom{Rodr\'{\i}guez-G\'omez}

\def\Z{\ensuremath{\mathbb{Z}}}

\def\R{\ensuremath{\mathbb{R}}}

\def\mn{{\mu\nu}}

\def\makeatletter{\catcode`\@=11}% 11:letter
\makeatletter
\def\mathbox#1{\hbox{$\m@th#1$}}%
\def\math@ccstyles#1#2#3#4#5#6#7{{\leavevmode
      \setbox0\mathbox{#6#7}%
      \setbox2\mathbox{#4#5}%
      \dimen@ #3%
      \baselineskip\z@\lineskiplimit#1\lineskip\z@
      \vbox{\ialign{##\crcr
             \hfil \kern #2\box2 \hfil\crcr
             \noalign{\kern\dimen@}%
             \hfil\box0\hfil\crcr}}}}
\def\mathaccstyles{\math@ccstyles\maxdimen}
\def\maththroughstyles{\math@ccstyles{-\maxdimen}}
\def\unity%
 {\maththroughstyles{.45\ht0}\z@\displaystyle {\mathchar"006C}\displaystyle 1} 
\begin{document}

\rightline{UG-FT-216/07}
\rightline{CAFPE-86/07}
\rightline{FFUOV-07/03}
\rightline{PUPT-2231}
%\rightline{hep-th/yymmnnn}
\rightline{April 2007}
\vspace{1.5truecm}

%%%%%%%%%%%
\centerline{\huge \bf A KK-monopole giant graviton in $AdS_5\times Y_5$}
%\vspace{.5cm}
%\centerline{\huge \bf configuration in $AdS_5\times Y_5$  }
\vspace{1.3truecm}

\centerline{
    {\large \bf Bert Janssen${}^{a,}$}\footnote{E-mail address: 
                                  {\tt bjanssen@ugr.es}},
    {\large \bf Yolanda Lozano${}^{b,}$}\footnote{E-mail address:
                                  {\tt
				    yolanda@string1.ciencias.uniovi.es}}
    {\bf and} 
    {\large \bf Diego Rodr\'{\i}guez-G\'omez${}^{c,}$}\footnote{E-mail address: 
                                  {\tt drodrigu@princeton.edu}}
                                                            }
\vspace{.6cm}
\centerline{{\it ${}^a$ Departamento de F\'{\i}sica Te\'orica y del Cosmos and}}
\centerline{{\it Centro Andaluz de F\'{\i}sica de Part\'{\i}culas Elementales}}
\centerline{{\it Universidad de Granada, 18071 Granada, Spain}}

\vspace{.4cm}
\centerline{{\it ${}^b$ Departamento de F{\'\i}sica,  Universidad de Oviedo,}}
\centerline{{\it Avda.~Calvo Sotelo 18, 33007 Oviedo, Spain}}

\vspace{.4cm}
\centerline{{\it ${}^c$ Department of Physics, Princeton University,}}
\centerline{{\it Princeton, NJ 08540, U.S.A.}} 

\vspace{2truecm}

%%%%%%%
\centerline{\bf ABSTRACT}
\vspace{.5truecm}

\noindent
We construct a new giant graviton solution in $AdS_5\times Y_5$, with $Y_5$ a quasi-regular 
Sasaki-Einstein manifold, consisting on a Kaluza-Klein monopole wrapped around the $Y_5$ and with 
its Taub-NUT direction in $AdS_5$. We find that this configuration has minimal energy when put 
in the centre of  $AdS_5$, where it behaves as a massless particle. When we take $Y_5$ to be $S^5$,
we provide a microscopical description in terms of multiple gravitational waves expanding into 
the fuzzy $S^5$ defined as an $S^1$ bundle over the fuzzy $CP^2$. Finally we provide a possible 
field theory dual interpretation of the construction.

%%end of title page
%%%%%%%%%%%%%%%%%%%%%%%%%%%%%%%%%%%%%%%%%%%%%%%%%%%%%%%%%%%%%
%%%%%%%%%%%%%%%%%%%%%%%%%%%%%%%%%%%%%%%%%%%%%%%%%%%%%%%%%%%%%
\newpage
\sect{Introduction}

As it is well-known, giant gravitons are stable brane configurations with non-zero angular momentum, 
that are wrapped on $(n-2)$- or $(m-2)$-spheres in $AdS_m\times S^n$ spacetimes and carry a dipole 
moment with respect to the background gauge potential \cite{GST,GMT,HHI,DJM}. They are not 
to\-po\-lo\-gi\-cally stable, but are at dynamical equilibrium because the contraction due to the 
tension of the brane is precisely cancelled by the expansion due to the coupling of the angular 
momentum to the background flux field. These spherical brane configurations turn out to be massless, 
conserve the same number of supersymmetries and carry the same quantum numbers of a graviton.

Giant graviton configurations were first proposed as a way to satisfy the stringy exclusion 
principle implied by the AdS/CFT correspondence \cite{GST}. The spherical $(n-2)$-brane expands 
into the $S^n$ part of the geometry with a radius proportional to its angular momentum. Since this 
radius is bounded by the radius of the $S^n$, the configuration has associated a maximum angular 
momentum. The $(m-2)$-brane giant graviton configurations \cite{GMT,HHI}, on the other hand, expand 
into the $AdS_m$ part of the geometry, and they do not satisfy the stringy exclusion principle.  
For a discussion on the degeneracy of these two types of giant gravitons and 
the point-like graviton, we refer for instance to \cite{BS} and references therein.

The construction of giant gravitons has also been generalised to $AdS_5\times Y_5$ spacetimes, where 
$Y_5$ is a Sasaki-Einstein manifold. In \cite{MS} and \cite{BM} a D3-brane 
wrapped around the angular $S^3$ of the $AdS_5$ and moving along the Reeb vector of the 
Sasaki-Einstein space was considered, yielding a dual giant graviton. 
A generalisation of giant graviton configurations preserving 1/4 or 1/8 of the supersymmetries has also been considered by Mikhailov  \cite{Mikhailov} in the $AdS_5 \times T^{1,1}$ spacetime.

In this paper we find a new giant graviton configuration in $AdS_5\times Y_5$, which consists of 
a Kaluza-Klein (KK) monopole with internal angular momentum, wrapping the $Y_5$ part of the geometry 
and with Taub-NUT direction in the $AdS_5$ part. This solution has distinguishing features 
with respect to the previous giant graviton solutions constructed in the literature. First, 
the monopole does not couple to the 4-form potential of the background and the configuration is 
therefore not at a dynamical equilibrium position. Still it is stable because by construction it 
is wrapped around the entire $Y_5$. Secondly,  it has a fixed size $L$, the ``radius'' of the $Y_5$, 
independent of the momentum of the configuration. In fact the energy of the monopole depends only on 
its position in the $AdS_5$ part of the spacetime, and it is minimised when the monopole sits at the 
centre of $AdS_5$, where it behaves as a massless particle. In this sense this new giant graviton 
configuration does not provide a realisation of the stringy exclusion principle. However, its mere 
existence is sufficiently surprising to motivate a closer look at the configuration. Furthermore, 
the fact that the giant graviton is built up from a Kaluza-Klein monopole, could lead to interesting 
view points in the context of the AdS/CFT correspondence.

The organisation of this paper is as follows. In section 2 we present the Kaluza-Klein monopole 
giant graviton solution. We start by introducing our probe monopole and then construct an action 
suitable to describe it. We then calculate the energy of the configuration and show that when the 
monopole sits at the centre of $AdS_5$ it behaves as a massless particle. In section 3 we move to 
consider the microscopical description of this configuration in terms of expanding gravitational 
waves. Given that the fuzzy version of an arbitrary Sasaki-Einstein manifold is not known, we 
particularise to the case in
which $Y_5=S^5$. The fuzzy 5-sphere that we consider is defined as an $S^1$ bundle over the fuzzy 
$CP^2$. This fuzzy manifold has been successfully used in the microscopical description of 
5-sphere giant gravitons  \cite{JLR3, LR1}, and of the baryon vertex with magnetic flux \cite{JLR4}. 
In these examples the fibre structure of the $S^5$ plays a crucial role in the construction.
Finally in section 4 we present a candidate description of our configuration in the field theory 
side. We end with some conclusions in section 5.

%%%%%%%%%%%%%%%%%%%%%%%%%%%%%%%%%%%%%%%%%%%%%%%%%%%%%%%%%%%%%%%%%%%%%%%%%%%%%%

\section{A new giant graviton solution}

\subsection{The Kaluza-Klein monopole probe}

Consider the $AdS_5 \times Y_5$ spacetime, with $Y_5$ a quasi-regular five-dimensional 
Sasaki-Einstein manifold. All these Sasaki-Einstein manifolds have a constant norm Killing 
vector, called the Reeb vector. For the cases we are interested in, the $U(1)$ action of the 
Reeb vector is free and the quotient space is (at least locally) a four-dimensional regular
 K\"ahler-Einstein manifold $M_4$ with positive curvature. In that case the metric on $Y_5$ 
can (at least locally) be written as a $U(1)$ fibre bundle over the $M_4$,
\be
ds^2_Y = ds^2_{M} + (d\psi + B)^2,
\ee
where $ds^2_M$ is the metric on the $M_4$ and  the Killing vector $k^\mu = \delta^\mu_\psi$ is 
the Reeb vector. The K\"ahler form on $M_4$ is related to the fibre connection $B$ 
via $\omega_M = \frac{1}{2}dB$.\footnote{For an extensive summary on the properties of Sasaki-Einstein  
    manifolds we refer to \cite{MSY}.} The $AdS_5 \times Y_5$
background contains as well a non-vanishing 4-form RR-potential.

Using the $U(1)$ decomposition above, the metric of $AdS_5 \x Y_5$ can be written as
\bea
ds^2 \ = \ -(1+ \frac{r^2}{L^2})dt^2 \ + \ \frac{dr^2}{(1+ \frac{r^2}{L^2})}
        \ + \ \frac{r^2}{4} \Bigl[ d\Omega_2^2+ (d\chi + A)^2 \Bigr] 
        + \ L^2 \Bigr[ ds^2_{M} + (d\psi + B)^2 \Bigr],
\eea
where we have used global coordinates in the $AdS$ part and written the angular $S^3$ contained in 
$AdS_5$ as a $U(1)$ fibre over $S^2$.
$A$ and $B$ stand for the connections of the $S^3$ and $Y_5$ fibre bundles respectively.
In these coordinates, the fibre directions $\chi$ and $\psi$ are clearly globally defined 
isometry directions.

Consider now a KK-monopole wrapped on the $Y_5$, with Taub-NUT direction $\chi$ and
propagating along $\psi$. This will be our KK-monopole probe.
In order to study the dynamics of this monopole we start by constructing an action suitable to 
describe it.

The effective action describing the
dynamics of the Type IIB Kaluza-Klein monopole was constructed in \cite{EJL}. Like 
the Type IIA NS5-brane, to which it is related by T-duality along the Taub-NUT direction, the 
action for the monopole is described by a six-dimensional $(2,0)$ tensor supermultiplet, which 
contains a self-dual 2-form $ \hat W^+_{ab}$ and 5 scalars $\{X^i, \omega, \tilde\omega \}$. 
The self-dual 2-form is associated to the (S-duality invariant) configuration of the monopole 
intersecting a D3-brane, wrapped on the Taub-NUT direction. The worldvolume scalars $\omega$ 
and $\tilde\omega$ are associated with the intersections of D5- and NS5-branes respectively, and 
form a doublet under S-duality. Finally the scalars $X^i$ (with $i=1,2,3$) are the embedding 
scalars, that describe the position of the monopole in the transverse space. Note that although 
the worldvolume of the monopole is six-dimensional, its position is specified by only three 
embedding scalars. This is because the Taub-NUT direction is considered to be transverse, but 
being an isometry direction it does not yield a dynamical degree of freedom. The KK-monopole 
action takes in fact the form of a gauged sigma model, where the degree of freedom 
corresponding to the Taub-NUT direction is gauged away \cite{BJO}. Due to the presence of the 
self-dual two-form $ \hat W^+_{ab}$, there is no straightforward covariant formulation of 
the action (see for example \cite{PST}). However, like in the case of the five-brane \cite{BRO, BLO}, 
it is possible to give an approximation, expanding the action to quadratic order in the self-dual 
two-form.

In our case, the situation is actually simpler. As our KK-monopole probe is wrapped 
around $Y_5$, the $U(1)$ fibre direction $\psi$ is contained in its worldvolume. 
Therefore it is possible to effectively compactify the monopole over the fibre direction and to
consider the (much simpler) action for a wrapped KK-monopole. Moreover, momentum charge along this 
$U(1)$ fibre direction can easily be induced by switching on an appropriate magnetic flux in the 
worldvolume.

The field content of the wrapped monopole is given by the five-dimensional $(1,1)$ vector multiplet, 
which contains 5 scalars and one vector, and is the dimensional reduction of the six-dimensional 
$(2, 0)$ tensor multiplet. The self-duality condition becomes a Hodge-duality condition 
between the vector and a two-form, which does not appear 
explicitly in the action.\footnote{This is very similar to the M5-brane case. The unwrapped M5-brane 
   contains a self-dual 2-form in its worldvolume, whereas the M5-brane wrapped on the eleventh 
   direction (the D4) depends on an unconstrained five-dimensional vector field.}
In this way an action can be constructed to all orders 
in the field strength. In practise, the action of the wrapped monopole is most easily constructed from 
the action of the Type IIA KK-monopole, as the latter has the six-dimensional $(1,1)$ vector multiplet 
as its worldvolume field content \cite{BEL}. After T-dualising along a worldvolume direction, the 
resulting action describes a Type IIB monopole wrapped along the T-duality direction, with an 
effectively five-dimensional worldvolume. As our KK-monopole probe is wrapped on the $S^1$ fibre 
direction of the $Y_5$, its spatial worldvolume  becomes effectively $\R \times M_4$. 

%%%%%%%%%%%%%%%%
\subsection{The action for the wrapped monopole}
 
The starting point is the action for the Type IIA Kaluza-Klein monopole constructed in 
\cite{BEL}, which we compactify along a worldvolume direction and T-dualize. 
The resulting action describes 
a Type IIB Kaluza-Klein monopole which is wrapped on the T-duality direction and has, 
effectively, a five-dimensional worldvolume. The T-duality direction appears in the action 
as a new isometric direction, whose Killing vector we denote by $k^\mu$. On the other hand 
the Killing vector associated with the Taub-NUT direction is denoted by $\ell^\mu$. The explicit 
action is given by
\begin{eqnarray}
\label{IIBw}
S &=& -T_4 \int d^5\sigma \, e^{-2\phi} k \ell^2\, 
         \sqrt{|\det (D_a X^\mu D_b X^\nu g_{\mu\nu} +  e^\phi k^{-1} \ell^{-1} {\cal F}_{ab})|} 
                                                            \nonumber\\
  && - T_4 \int d^5\sigma \Bigl\{P[\incl_k \incl_\ell
         N^{(7)}]-P\Bigl[\incl_\ell C^{(4)}\Bigr] \wedge {\cal F}
                -\frac12 P\Bigl[\frac{k^{(1)}}{k^2}\Bigr] \wedge {\cal F}\wedge {\cal F}
                  + \dots \Bigr\},
\end{eqnarray}
where the scalars $k$ and $\ell$ are the norm of $k^\mu$ and $\ell^\mu$ respectively, $k^{(1)}$ 
denotes the 1-form with components $k_\mu$ and $(\incl_\ell \incl_k \Omega)_{\mu_1 ... \mu_n} = 
\ell^\rho k^\nu \Omega_{\nu\rho\mu_1 ... \mu_n}$. In this action the pull-backs into the worldvolume 
are taken with gauge covariant derivatives
\begin{equation}
D_a X^\mu=\partial_a X^\mu - k^{-2} k_\nu \partial_a X^\nu k^\mu 
                           - \ell^{-2} \ell_\nu \partial_a X^\nu \ell^\mu,
\end{equation}
which ensure local invariance under the isometric transformations generated by the two Killing 
vectors
\begin{equation}
\delta X^\mu=\Lambda^{(1)}(\sigma)k^\mu+\Lambda^{(2)}(\sigma)\ell^\mu\, .
\end{equation}
In this way the embedding scalars corresponding to the isometry directions are eliminated as 
dynamical degrees of freedom and the action is given by a gauged sigma model of the type 
first considered in \cite{BJO}. 

The two-form field strength ${\cal F}$ is defined as
\begin{equation}
\label{efe}
{\cal F}=2\partial V^{(1)}+  P[\incl_k \incl_\ell C^{(4)}]\, ,
\end{equation}
where the worldvolume vector field $V^{(1)}$ is the T-dual of the vector field of the Type IIA 
monopole (or, alternatively, the dimensional reduction of the self-dual two-form $\hat W^+$). 
While in the Type IIA monopole the vector field is associated to D2-branes wrapped on the Taub-NUT 
direction, in the IIB case it is associated to D3-branes, wrapped on both Killing directions.

The action of the Type IIA monopole contains as well a worldvolume scalar associated to strings 
wrapped on the Taub-NUT direction. This field gives, upon T-duality, a worldvolume scalar $\omega$
which forms a doublet under S-duality with the T-dual $\tilde \omega$ of the component of the IIA 
vector field along the 
T-duality direction. These two scalars are necessary in order to compensate for the two 
degrees of freedom associated to the two transverse scalars that have been eliminated from 
the action through the gauging procedure. This scalar doublet does however not play a role in 
our construction and has therefore been set to zero in our action above. The action (\ref{IIBw}) 
should then be regarded as a truncated action suitable for the study of the 
wrapped monopole in the $AdS_5 \times Y_5$ background.

In the Chern-Simons part of the action we find a coupling to $N^{(7)}$, the tensor field dual to 
the Taub-NUT Killing vector $\ell_\mu$, considered as a 1-form. The contraction 
$i_\ell N^{(7)}$ is the field to which  
a KK-monopole with Taub-NUT direction $\ell^\mu$ couples minimally (see \cite{BEL}). In (\ref{IIBw}) 
this field is further contracted with the second isometric direction $k^\mu$, indicating that 
the monopole is wrapped along this direction. More importantly for our construction below, the 
second coupling in the CS action involves the momentum operator $P[k^{(1)}/k^2]$, associated to the
isometric direction with Killing vector $k^\mu$. Therefore, it is possible to induce momentum 
charge in this isometric direction, with an appropriate choice of $\cF$. As we will show below,
we will make use of this coupling to let the monopole propagate along the isometric direction $\psi$.  
Finally, the dots indicate couplings to other Type IIB background fields which do not play a role in 
our construction.

%%%%%%%%%%%%%%%
\subsection{The giant graviton solution}
\label{solution}
  
Let us now particularise the action (\ref{IIBw}) to our probe KK-monopole.  We take our monopole 
wrapped on the transverse $Y_5$. Therefore the fibre direction of the decomposition of $Y_5$  as a 
$U(1)$ fibre bundle over $M_4$ is identified as the isometric worldvolume direction in (\ref{IIBw}), 
and $M_4$ as the effective four-dimensional spatial worldvolume.
The Taub-NUT direction is taken along 
the $S^1$ fibre direction of the $S^3$ contained in $AdS_5$. Therefore, we have explicitly
\begin{equation}
k^\mu=\delta^\mu_\psi\, , \hspace{2cm}  \ell^\mu=\delta^\mu_{{\chi}}.
\label{isometries}
\end{equation} 
With this choice of Killing directions the contribution of the 4-form
RR-potential of the $AdS_5\times Y_5$ background to the action
vanishes. This is so because both couplings to $C^{(4)}$, in
(\ref{efe}) and in (\ref{IIBw}), involve directions along $Y_5$, plus
the Taub-NUT direction, $\chi$, which lives in the $AdS_5$ part of the spacetime.

Furthermore, in order to induce momentum charge in the $\psi$ direction we choose the worldvolume 
vector field $V$ proportional to the curvature tensor of the $Y_5$ fibre connection $B$, such that 
\be
\cF = {}^*\cF\, ,  \hsp{2cm}
\int_M \cF\wedge \cF = 2 n^2 \Omega_M,
\label{fluxM}
\ee
where $\Omega_M$ is the volume of $M_4$ and the Hodge star is taken
with respect to the metric on this manifold\footnote{The integral above is 
      non-zero because it is the product of two integrals, $\oint {\cal F}$, over non-trivial 
      two-cycles in $M_4$ (see for example \cite{GMSW}). Since $\oint {\cal F}=2\pi n$ due to 
      Dirac quantisation condition, $n$ represents the winding number of D3-branes wrapped around 
      each of the two cycles. For our construction we have chosen the same winding number in both 
      cycles in order to preserve the self-duality condition
      (\ref{fluxM}). }. With this Ansatz ${\cal F}$ satisfies trivially
the Bianchi identities. 
Then, through the second coupling in the Chern-Simons part of the action (\ref{IIBw}), we have that
\begin{equation}
\frac{T_4}{2}\int_{\R \x M_4} P\Bigl[k^{-2}k^{(1)}\Bigr]\wedge \cF\wedge \cF 
\ = \  n^2 T_W \int dt\ P\Bigl[k^{-2}k^{(1)}\Bigr]\, ,
 \end{equation}
where we have used the fact that the tension of the wrapped monopole is related to the tension of the
point-like object carrying momentum charge (the gravitational wave) through $\Omega_M T_4=T_W$.
Therefore, with this Ansatz for $\cF$, we are dissolving in the worldvolume $n^2$ momentum charges in 
the $\psi$ direction. Notice that the instantonic nature of
(\ref{fluxM}) guarantees that the equations of motion for ${\cal F}$
are satisfied.

A second remarkable property of the Ansatz (\ref{fluxM}) is that the determinant of 
$(P[g]_{ab} + \cF_{ab})$ is a perfect square \cite{JLR3, JLR4},
such that the Born-Infeld 
part of the action (\ref{IIBw}) gives rise to
\begin{equation}
S =- T_4 \int dt \ d\Omega_M\  \frac{Lr^2}{4} \sqrt{\Bigl(1+ \frac{r^2}{L^2}\Bigr)
                     \Bigl[ \frac{L^4}{8} + \frac{4n^2}{L^2 r^2}\Bigr]^2 |g_M|}, 
\end{equation}
which after integration over $M_4$ gives rise to the following Hamiltonian
\begin{equation}
\label{Emac}
H=\frac{n^2 T_W}{L}\ \sqrt{1+\frac{r^2}{L^2}}\ \Bigl[1+\frac{L^6 r^2}{32 n^2}\Bigr].
\end{equation}

The energy of the configuration is therefore a function of the radial coordinate $r$ of $AdS_5$ and 
is clearly minimised when $r=0$, that is, when the monopole is sitting at the centre of $AdS_5$. 
Moreover, for this value of $r$ the energy is given by
\begin{equation}
E=\frac{n^2 T_0}{L}=\frac{P_\psi}{L}\, .
\end{equation}

Therefore, the configuration that we have proposed behaves as a giant graviton: it has the energy  
of a massless particle with momentum $P_\psi$ but  clearly has some finite radius $L$, as it is 
wrapped around the entire $Y_5$. Since it saturates
a BPS bound it is a solution of the
equations of motion.
Finally, 
we should note that there is no dynamical 
e\-qui\-li\-brium between the brane tension and the angular momentum, as in the traditional 
giant graviton configurations 
of \cite{GST, GMT, HHI, DJM}. However the stability of the configuration is still guaranteed due to 
the fact that it wraps the entire transverse space.
 
%%%%%%%%%%%%%%%%%%%%%%%%%%%%%%%%%%%%%%%%%%%%%%%%%%%%%%%%%%%%%%%%%
\sect{A microscopical description in terms of dielectric gravitational waves}

It is by now well-known that the traditional giant graviton configurations of \cite{GST,GMT,HHI, DJM}
can be described microscopically in terms of multiple gravitational waves expanding into (a fuzzy 
version of) the corresponding spherical brane by Myers' dielectric effect \cite{Myers}. In particular, 
the M5-brane 
giant graviton configurations of the $AdS_4\times S^7$ and $AdS_7\times S^4$ spacetimes have been 
described in terms of multiple M-waves expanding into a fuzzy 5-sphere that is defined as an $S^1$ 
bundle over a fuzzy $CP^2$ \cite{JLR3}. A non-trivial check of the validity of this description is 
that it agrees exactly with the spherical brane description in \cite{GST,GMT} when the number of 
gravitons is very large.

In this spirit one would expect that the KK-monopole giant graviton configuration constructed in 
the previous section would be described microscopically in terms of dielectric Type IIB gravitational 
waves expanding into a fuzzy $Y_5$. The fuzzy version of general quasi-regular Sasaki-Einstein 
manifolds is however not known. Therefore we will restrict to the case in which $Y_5$ coincides 
with the 5-sphere. In this case we will see that the waves expand into a fuzzy 5-sphere, defined 
as an $S^1$ bundle over a fuzzy $CP^2$, i.e. the same type of fuzzy manifolds involved in the 
description used in \cite{JLR3} (see also \cite{LR1,JLR4}).

Consider now a number $N$ of coinciding gravitational waves in the $AdS_5\times S^5$ background. 
The action describing multiple gravitational waves in Type IIB was constructed in \cite{JL1,JLR1} 
and it contains the following
couplings:
\bea
S_W &=& - \ T_W \int d\tau \ \STr\Bigl\{ k^{-1} \sqrt{ 
     -P\Bigl[E_\mn - E_{\mu i} (Q^{-1} -\delta)^i{}_j E^{jk} E_{k\nu}\Bigr] \det Q\ }\ \ \Bigl\} 
     \label{action}        \\ [.2cm]
&& 
+ \ T_W \int d\tau 
~\STr \Bigl\{ -  P [k^{-2} k^{(1)}] \ - i P[(\incl_X\incl_X)\incl_\ell C^{(4)}]    
- \half P[(\incl_X\incl_X)^2\incl_\ell\incl_k N^{(7)}] +\dots\Bigl\}
\nonumber
\eea
where
\bea
&& E_\mn = \cG_\mn - k^{-1} \ell^{-1} (\incl_k \incl_\ell C^{(4)})_\mn,  \hsp{1cm}
 \cG_\mn = g_\mn - k^{-2} k_\mu k_\nu -\ell^{-2} \ell_\mu \ell_\nu, \nnw
&& Q^\mu{}_\nu = \delta^\mu_\nu + i k \ell \ [X^\mu, X^\rho] E_{\rho\nu}, \hsp{1,6cm}
((\incl_X \incl_X) \incl_\ell C_4)_\lambda = [X^\rho, X^\nu] \ell^\mu C_{\mu\nu\rho\lambda}. \\[.2cm]
&& (\incl_\ell \incl_k \Omega)_{\mu_1 ... \mu_n} = \ell^\rho k^\nu \Omega_{\nu\rho\mu_1 ... \mu_n},
\nonumber
\eea

This action is valid to describe waves propagating in  backgrounds which contain two isometric 
directions, parametrised in the action by the 
Killing vectors $k^\mu$ and $\ell^\mu$. The wave action is actually a 
gauged sigma model in which the embedding scalars associated to the Killing directions
 are projected out. The physical meaning of $k^\mu$ is 
that it corresponds to the propagation direction of the gravitational waves, while $\ell^\mu$ is 
an isometry direction inherited from the T-duality operation involved in the construction of the 
action  (see \cite{JL1} and \cite{JLR1} for more details). 

Although the only non-zero term in the Chern-Simons action in the $AdS_5 \x S^5$ background is the 
dipole coupling to $C^{(4)}$, it is worth calling the attention to the quadrupole coupling to 
$N^{(7)}$. Indeed, this coupling shows that the waves can expand via a quadrupole effect into a 
monopole with Taub-NUT direction parametrised by $l^\mu$ and further wrapped around the $k^\mu$ 
direction. This monopole will then act as the source of $\incl_\ell \incl_k N^{(7)}$.

Let us now use the action (\ref{action}) to describe microscopically the KK-monopole of the 
previous section in the $AdS_5\times S^5$ background. In this case $M_4=CP^2$ and $ds^2_M$ 
stands for the Fubini-Study metric on the $CP^2$ (see for instance
\cite{Pope}). 
The identification of the isometry
directions of (\ref{action}) is then obvious. In order to account for the momentum in the 
worldvolume of the monopole, we identify the propagation direction of the waves with the $S^5$ 
fibre direction $\psi$, while the extra isometry will be identified with the Taub-NUT direction 
$\chi$ of the monopole:
\be
\label{killings}
k^\mu = \delta^\mu_\psi, \hsp{2cm} \ell^\mu = \delta^\mu_{\chi}. 
\ee 
With this choice of Killing vectors it is clear that the contribution
of $C^{(4)}$ to the action vanishes, both in the BI and in the CS
parts. Therefore, any dielectric effect will be purely gravitational
\cite{gdie1,gdie2,gdie3}.

Furthermore, we will have the gravitational waves expand into the entire five-sphere, whose fuzzy 
version we choose to describe as an $S^1$ bundle over the fuzzy $CP^2$. Therefore, we take the 
non-commutative scalars in (\ref{action}) to parametrise the fuzzy $CP^2$ base of the $S^5$. 

The fuzzy $CP^2$ has been extensively studied in the literature. For its use in the giant graviton 
context we refer to \cite{JLR3,LR1}, where more details on the construction that we sketch below 
can be found. $CP^2$ is the coset manifold $SU(3)/U(2)$ and can be defined as the 
submanifold of $\R^8$ determined by the constraints
\be
\label{condi}
\sum_{i=1}^8 x^i x^i=1\ , 
\hspace{2cm}
\sum_{j,k=1}^8d^{ijk}x^j x^k =\frac{1}{\sqrt{3}}x^i\ ,
\ee
where $d^{ijk}$ are the components of the totally symmetric $SU(3)$-invariant tensor. A  fuzzy 
version of $CP^2$ can  then be obtained by imposing the conditions (\ref{condi}) at the level of 
matrices (see for example \cite{ABIY}). We define a set of coordinates $X^i$ $(i=1,\dots,8)$ as
\begin{equation}
\label{defX}
X^i=\frac{T^i}{\sqrt{(2N-2)/3}},
\label{X(T)}
\end{equation}
where $T^i$ are the $SU(3)$ generators in the $N$-dimensional irreducible representations $(k,0)$ or 
$(0,k)$, with $N=(k+1)(k+2)/2$. The first constraint in (\ref{condi}) is  then trivially
satisfied through the quadratic Casimir $(2N-2)/3$ of the group, whereas the rest of the constraints 
are satisfied for any $N$ (see \cite{ABIY, JLR3} for the details).
The commutation relations between the $X^i$ are given by
\be
[X^i, X^j] = \frac{i  \ f^{ijk}}{\sqrt{(2N-2)/3}} X^k,
\label{[XX]}
\ee
with $f^{ijk}$ the structure constants of $SU(3)$ in the algebra of the Gell-Mann matrices 
$[\lambda^i,\lambda^j]=2i f^{ijk} \lambda^k$.

Substituting the Ans\"atze (\ref{X(T)}) and (\ref{killings}) in the action (\ref{action}), we find
\be
S = -T_W \int d\tau \ \Str\Bigl\{ L^{-1}
  \sqrt{\Bigl(1 + \frac{r^2}{L^2} \Bigr)
           \Bigl[ \unity  + \frac{3L^6 r^2}{32(N-1)} X^2 \Bigr]^2 }\ \ \Bigr\} ,
\label{action2}
\ee
up to order $N^{-2}$. Here we have dropped those contributions to
${\det}\ Q$ that vanish when taking the symmetrised trace, and
ignored higher powers of $N$ which will vanish in the large $N$
limit.\footnote{These terms cannot be nicely arranged into higher
  powers of the quadratic Casimir without explicit use of the constraints.}

Taking the symmetrised trace we arrive at the following Hamiltonian
\be
H=  \frac{N T_W}{L} \ \sqrt{1 + \frac{r^2}{L^2}}\ \  \Bigl[ 1 + \frac{L^6 r^2}{32(N-1)}\Bigr],
\label{offshell}
\ee
which, in the large $N$ limit, is in perfect agreement with the Hamiltonian for the spherical 
KK-monopole, given by (\ref{Emac}). To see this we should recall that
in the macroscopical description the momentum charge of the
configuration is the number of waves dissolved in the worldvolume, and
is therefore given by $n^2$. In the microscopical description 
the momentum charge is given directly by the number of microscopic
waves, $N$. Therefore in the large $N$ limit $N$ and $n^2$ must coincide.
The
Hamiltonian in (\ref{offshell}) is also a 
function of the radial coordinate $r$ of $AdS_5$ and it is minimised at $r=0$ where it takes 
the value $E=P_\psi/L$, thus corresponding to a giant graviton configuration.
 
%%%%%%%%%%%%%%%%%%%%%%%%%%%%%%%%%%%%%%%%%%%%%%%%%%%%
\sect{A possible interpretation in the dual field theory}

In this section we try to give a possible field theoretic interpretation of the giant 
graviton configuration that we have studied, along the lines of
\cite{HHI} (see also \cite{FT1, FT2, Pirrone, FT3}). We will discuss the
giant graviton configuration in the $AdS_5\times S^5$ background, but
we will later 
speculate on a possible generalisation to other Sasaki-Einstein spaces.

We have learnt in the previous sections that 
the fibre direction in
the $S^3$ 
contained in $AdS_5$ plays a crucial role in the construction of the
giant 
graviton configuration, as it is identified with the Taub-NUT
direction of 
the monopole. Therefore, it is useful to work in the global patch for $AdS$.

It is then natural to consider the dual field theory as living in $\R \x S^3$, where there is a 
conformal coupling to the curvature. The bosonic piece of the action reads 
\begin{equation}
S=\frac{1}{2}\int dt\ d\Omega_3 \ \Tr\Big\{
         \partial_{\mu}\Phi_a^* \partial^{\mu}\Phi_a 
        \ + \ \frac{1}{L^2}\Phi_a^* \Phi_a 
        \ + \ \frac{1}{4g^2}[\Phi_a,\Phi_b^*]^2\Big\}\ ,
\label{actionS3}
\end{equation}
where $L$ is the radius of the $S^3$ and
the  $\Phi_a$ (with $a=1,2,3$) are the complexification of the 6
adjoint real scalars 
$X^i$ of $\mathcal{N}=4$ SYM. After defining
$\Phi_a=X^a+iX^{a+3}$, 
only an $SU(3)$ subgroup of the original $SO(6)$ R-symmetry group remains explicit.

Regarding  $S^3$ as an $S^1$ bundle over $S^2$ it seems a consistent
truncation to  assume that the $\Phi_a$ do 
not depend on the fibre coordinate. Actually, this will be 
the field theory analogue of the fact that this direction corresponds 
to the Taub-NUT direction of the monopole on the gravity side. Taking
adapted coordinates to the $U(1)$ fibration we have
\begin{equation}
S=2 \pi \int dt\int d\Omega_2 \ \Tr\Big\{
        - \partial_t\Phi_a^*\partial_t\Phi_a
       \ - \ 4\Phi_a^* \Delta_{S^2}\Phi_a
       \ + \ \frac{1}{L^2}\Phi_a^*\Phi_a
       \ + \ \frac{1}{4g^2}[\Phi_a,\Phi_b^*]^2\Big\},
\label{scalaraction}
\end{equation}
where $\Delta_{S^2}$ is the Laplacian in the 2-sphere.

We can then expand the scalars in spherical harmonics $\Phi_a^{(lm)}$ on the two-sphere. Given 
that we will be interested in the lowest energy modes, we will truncate all of them except 
the massless mode $\Phi_a^{(0)}$, which corresponds to the constant mode on the $S^2$. Furthermore, 
we consider the following Ansatz for the gauge and $SU(3)$ dependence of our fields
\begin{equation} 
\Phi_a^{(0)}=e^{if(t)}\mathcal{M}_a\otimes J_a,
\label{scalaransatz}
\end{equation}
where $f(t)$ is an arbitrary function of time, the traceless matrix $\mathcal{M}_a$ is defined as
\begin{equation}
\mathcal{M}_a= {\rm diag}\ \Bigl(v_a,
-\frac{v_a}{M-1},\cdots,-\frac{v_a}{M-1} \Bigr)\, ,
\end{equation}
and the $J_a$'s are $SU(2)$-generators in a $k$-dimensional representation. Since the total rank of 
the gauge group is $N$ we should have that $kM=N$. Indeed, in this branch the gauge group breaks 
into $SU(M)$. The gauge transformations which are left are those of the form 
\begin{equation}
\Phi_a\rightarrow g\Phi_a g^\dagger,\hsp{1cm} 
g={\rm diag}(\tilde{g}_{k},\cdots,\tilde{g}_{k})\ ,
\end{equation}
where the $\tilde{g}_{k_i}$ are $SU(2)$ gauge transformations of dimension $k$.

The action then reduces to
\begin{equation}
S= 8\pi^2 \frac{C_2(k)M}{M-1}\int dt \Big[ - (f')^2  v^2 
              \ +\ \frac{ v^2}{L^2}\Big]\ ,
\label{actie2}
\end{equation}
where $v^2= \delta^{ab}v_a v_b $ is to be interpreted as a non-dynamical parameter whose value will
determine the minima of the potential. In addition, $C_2(k)$ is the Casimir of the $SU(2)$ 
$k$-dimensional representation.
In this action $f(t)$ is a cyclic variable and therefore its conjugate
momentum, $p$, will be conserved. The Hamiltonian is given by
\begin{equation}
H=p^2\frac{(M-1)}{32\pi^2 v^2 C_2(k) M} + \frac{8\pi^2MC_2(k) v^2}{(M-1)L^2},
\end{equation}
which has a minimum for $v^2 = \frac{(M-1)}{32\pi^2C_2(k)M}pL$. Remarkably the on-shell energy is 
precisely the dispersion relation 
\begin{equation}
E=\frac{p}{L}.
\end{equation}
Therefore, the configuration (\ref{scalaransatz}) can be seen as a
 massless particle. Furthermore, out of the original $SU(3)$ rotating
 our $\Phi_a$ just an $SU(2)$ survives, given that with our Ansatz the 
$\Phi_a$ become a vector of $SU(2)$. Thus, the moduli space reduces
to $SU(3)/U(2)$, which is precisely the symmetry of $C P^2$ as a
coset space, which is in turn the manifold wrapped by our KK-monopole.

Given that our construction of the wrapped KK-monopole works not
just in the $S^5$ case, but also in more generic spaces, we expect a similar field theory 
description for the dual of a generic Sasaki-Einstein space. In supporting 
this claim, let us notice that the potential term did not play any
role in the $S^5$ case, because with the Ansatz we assumed, it vanishes. 
In the generic case, we will assume a similar Ansatz for our fields, namely
\begin{equation}
X_{\alpha}=e^{if(t)}\mathcal{M}\otimes G_{\alpha}\ ,
\end{equation}
where now for simplicity we take the same $\mathcal{M}$
matrix as before but with all 
the $v$'s identical. The $G_{\alpha}$ are the generators of the global symmetry group $G$. 
Given this Ansatz, we expect that the superpotential does not play any role,
 not even in the most generic $Y^{p,q}$ case. 
In addition, since we take $AdS$ in the
global patch, the field theory will be defined in
$\R\times S^3$, so we will always
have the conformal coupling to the curvature. 
Just this term, together with the kinetic energy, is enough to
reproduce a dispersion relation of the form $E\sim p$. 

In the general case we can also regard the
$S^3$ as an $S^1$ bundle over $S^2$, and take our fields independent of the 
$U(1)$. This is the field theory counterpart of the presence of the Taub-NUT 
direction in the gravity side. In addition, out of the full global 
symmetry group $G$ we will just keep the subgroup $g$
compatible with our Ansatz,\footnote{Note that $g$ may involve discrete subgroups such as $\Z_k$} 
so we would expect a moduli space of the form $G/g$. Let us consider for example the conifold. In
this case the global symmetry group is $SU(2)\times SU(2)$. Therefore
we have to first reduce it to the diagonal $SU(2)$ and then take the conifold scalars $A$ and
$B$ to be $e^{if(t)}\mathcal{M}$. This leaves an $[SU(2)\times SU(2)]_D/U(1)$ 
moduli space, which is the symmetry of 2 2-spheres, and this is in
turn what one would get from the gravity side.

Finally, we would like to note that our construction is quite generic. {}From the gravity point 
of view we just require that the momentum of
the KK-monopole wrapping the five-dimensional space is taken along a $U(1)$
fibre direction, and that its Taub-NUT direction is along the $S^1$ in the
decomposition of the $S^3\subset AdS_5$ as an $S^1$ bundle over $S^2$.
In the field theory side our requirements are also quite generic.
The existence of the Taub-NUT direction is reflected on the fact that
the SCFT is defined in  $\R\times S^3$ and  $S^3$ is taken as $S^1$
 over $S^2$. 
In addition, our description is not sensitive to the superpotential, 
which we believe is the counterpart to the fact that the KK-monopole 
wraps the whole 5-dimensional manifold. Then, we are left with the kinetic term
 and 
the conformal coupling to curvature, which is enough to ensure the
right
dispersion relation. Since in general our Ansatz will reduce the
 original 
global symmetry, the moduli space will be $G/g$, which we believe will 
correspond in general to the symmetry of the 4-dimensional base which the KK-monopole wraps.

%%%%%%%%%%%%%%%%%%%%%%%%%%%%%%%%%%%%%%%%%%%%%%%%%%%%%%%%%%%%%%%%%%
\sect{Conclusions}

In this letter we have constructed a new type of giant graviton solution in $AdS_5\times Y_5$, with 
$Y_5$ a quasi-regular Sasaki-Einstein manifold. This solution consists on a Kaluza-Klein monopole 
with internal momentum, wrapped around the entire $Y_5$ and with Taub-NUT direction along the 
$AdS_5$ part. 

Although the dynamics of this monopole can be described using the effective action for the Type IIB 
Kaluza-Klein monopole constructed in \cite{EJL}, this action is only known to quadratic order in the 
self-dual 2-form of its six-dimensional $(2,0)$ tensor multiplet field content. However, given that 
$Y_5$ can be decomposed as a $U(1)$ bundle over a four-dimensional K\"ahler-Einstein manifold $M_4$, 
it is possible to use the action for a monopole wrapped on a $U(1)$ direction to describe it. This 
action, having the field content of the five-dimensional $(1,1)$ vector multiplet, is known to all 
orders. Moreover, it is possible to induce momentum charge along the $U(1)$ direction through a 
suitably chosen worldvolume vector field with non-zero instanton number. Using the action for a 
$U(1)$ wrapped monopole we have shown that the energy of the configuration depends on its radial 
position in the $AdS$ space and behaves as a massless particle when put in the origin, 
while having the size of the $Y_5$.

Given that the spherical monopole carries a non-vanishing momentum charge there should be a 
microscopical description of the same configuration in terms of expanding gravitational waves. 
This description would involve however the fuzzy version of $Y_5$, which is not known in general.
Therefore, we have restricted to the case in which $Y_5=S^5$, and let the multiple 
dielectric gravitational waves expand into a fuzzy 5-sphere. The fuzzy 5-sphere built up 
by the gravitational waves is constructed as an Abelian fibre bundle over a fuzzy $CP^2$, a 
construction that has been used before in the study of the traditional giant gravitons and the 
baryon vertex with magnetic flux. The configuration thus obtained turns out to exactly agree in the 
limit of large number of waves with the effective KK-monopole description.

We believe there are several reasons why this new giant graviton solution has not been found earlier 
in the literature. First of all, since it has no relation with the stringy exclusion principle it is 
not straightforward to find the corresponding state in the CFT
side.  Moreover, as we have shown in section 4, 
our scalar field configuration breaks the R-symmetry group in a rather peculiar way, making explicit 
the $U(1)$ fibre structure of the $S^3$. Secondly, the fact that it is built up from a Kaluza-Klein
monopole and not from a more ordinary type of brane, makes our construction more involved.
 
An interesting question to answer would be whether the KK-monopole giant graviton solution is 
supersymmetric or not. This is however difficult to answer, on the one hand due to the form of the 
Killing spinors in the particular coordinate system that we are using
and, on the other hand, due to the fact that the
 kappa-symmetry for the Kaluza-Klein monopole is not known. Yet, the fact that the configuration is 
massless implies that it saturates a BPS bound, which hints to the fact that it probably preserves 
some fraction of the supersymmetry. We would like to leave this problem for future investigations.

%%%%%%%%%%%%%%%%%%%%%%%%%%%%%%%%%%%%%%%%%%%%%%%%%%%%%%%%%%%%%
\vsp{1cm}

\noindent
{\bf Acknowledgements}\\
We wish to thank S.Benvenutti, B. Craps, D. van den Bleeken, T. van Proeyen and
A.V. Ramallo for the useful discussions.
The work of B.J.~is done as part of the program {\sl Ram\'on y Cajal} of the Ministerio de 
Educaci\'on y Ciencia (M.E.C.) of Spain. He is also partially supported by the M.E.C. under 
contract FIS 2004-06823 and by the Junta de Andaluc\'{\i}a group FQM 101. 
The work of Y.L. has been partially supported by the CICYT grant MEC-06/FPA2006-09199 (Spain) and by 
the European Commission FP6 program MRTN-CT-2004-005104, in which she is associated to Universidad
Aut\'onoma de Madrid. 
D.R.G. is supported by the Fullbright-M.E.C. fellowship FU-2006-0740.  
 
%%%%%%%%%%%%%%%%%%%%%%%%%%%%%%%%%%%%%%%%%%%%%%%%%%%%%%%%%%%%%%
%\newpage

%%%%%%%%%%%%%%%%%%%%%%%%%%%%%%%%%%%%%%%%%%%%%%%%%%%%%%%%%%%%%%

\begin{thebibliography}{99}

\bibitem{GST} J. McGreevy, L. Susskind, N. Toumbas, JHEP 0006 (2000) 008, hep-th/0003075.

\bibitem{GMT} M. Grisaru, R. Myers, \O. Tafjord, JHEP 0008 (2000) 040, hep-th/0008015.

\bibitem{HHI} A. Hashimoto, S. Hirano, N. Itzhaki, JHEP 0008 (2000) 051, hep-th/0008016.

\bibitem{DJM} S.R. Das, A. Jevicki, S.D. Mathur, Phys. Rev. D63 (2001) 044001, hep-th/0008088.

\bibitem{BS} I. Bena, D.J. Smith, Phys. Rev. D71 (2005) 025005, hep-th/0401173.

%\bibitem{BFHMS} S. Benvenuti, S. Franco, A. Hanany, D. Martelli, J. Sparks, 
%                JHEP 0506 (2005) 064,  hep-th/0411264.

\bibitem{MS} D. Martelli, J. Sparks, Nucl.Phys. B759 (2006) 292, hep-th/0608060.

\bibitem{BM} A. Basu, G. Mandal, {\it Dual Giant Gravitons in $AdS_m \times Y^n$ (Sasaki-Einstein)}, 
             hep-th/0608093.

\bibitem{Mikhailov} A. Mikhailov, JHEP 0011 (2000) 027, hep-th/0010206.

\bibitem{JLR3} B. Janssen, Y. Lozano, D. Rodr\'{\i}guez-G\'omez, 
               Nucl. Phys. B712 (2005) 371,  hep-th/0411181.

\bibitem{LR1} Y. Lozano, D. \RodrGom, JHEP 0508 (2005) 044, hep-th/0505073.

\bibitem{JLR4} B. Janssen, Y. Lozano, D. \RodrGom, JHEP 0611 (2006) 082, hep-th/0606264.

\bibitem{MSY} D. Martelli, J. Sparks, S.-T. Yau, {\it Sasaki-Einstein Manifolds and Volume 
              Minimisation}, hep-th/0603021.

\bibitem{EJL} E. Eyras, B. Janssen, Y. Lozano, Nucl. Phys. B531 (1998) 275, hep-th/9806169.

\bibitem{BJO} E. Bergshoeff, B. Janssen, T. Ort\'{\i}n, Phys. Lett. B410 (1997) 131, hep-th/9706117.

\bibitem{PST}  P. Pasti, D. Sorokin, M. Tonin, Phys. Lett. B398 (1997) 41, hep-th/9701037.

\bibitem{BRO} E. Bergshoeff, M. de Roo, T. Ort\'{\i}n, Phys. Lett. B386 (1996) 85, hep-th/9606118.

\bibitem{BLO} E. Bergshoeff, Y. Lozano, T. \Ortin, Nucl. Phys. B518 (1998) 363, hep-th/9712115.

\bibitem{BEL} E. Bergshoeff, E. Eyras, Y. Lozano, Phys. Lett. B430 (1998) 77, hep-th/9802199.

\bibitem{GMSW} J. Gauntlett, D. Martelli, J. Sparks, D. Waldram, JHEP 0506 (2005) 064 
                hep-th/0411264.

\bibitem{Myers} R. Myers,  JHEP 9912 (1999) 022, hep-th/9910053.

\bibitem{JL1} B. Janssen, Y. Lozano, Nucl. Phys. B643 (2002) 399,  hep-th/0205254.

\bibitem{JLR1} B. Janssen, Y. Lozano, D. \RodrGom, Nucl. Phys. B669 (2003) 363, hep-th/0303183.

\bibitem{Pope} C. Pope, Phys. Lett. B97 (1980) 417.

\bibitem{gdie1} 
V. Shakian, JHEP (2001)  0104, hep-th/0102200.


\bibitem{gdie2}
J. de Boer, E. Gimon, K. Schalm, J. Wijnhout, Annals Phys.313 (2004)
402, hep-th/0212250.

\bibitem{gdie3}
D. \RodrGom, JHEP 0601(2006) 079, hep-th/0509228.

\bibitem{ABIY} G. Alexanian, A. Balachandran, G. Immirzi, B. Ydri, 
               J. Geom. Phys. 42 (2002) 28, hep-th/0103023.

\bibitem{FT1} 
S. Corley, A. Jevicki, S. Ramgoolam, Adv. Theor. Math. Phys.(2002)
5:809-839, hep-th/0111222.

\bibitem{FT2}
D. Berenstein, JHEP 0407 (2004) 018, hep-th/0403110.

\bibitem{Pirrone} M. Pirrone, JHEP 0612 (2006) 064, hep-th/0609173.  

\bibitem{FT3}
E. Imeroni, A. Naqvi, JHEP 0703 (2007) 034, hep-th/0612032. 

\end{thebibliography}
\end{document}